\documentclass{article}

\usepackage{theorem}
\usepackage{latexsym}
\usepackage{amssymb}
\def\qed{\rule{1mm}{2.5mm}}

\theoremstyle{plain} \newtheorem{thm}{Theorem}[section]
\theoremstyle{plain} \newtheorem{prop}[thm]{Proposition}
\theoremstyle{plain} \newtheorem{lem}[thm]{Lemma}
\theoremstyle{plain} 
\theoremstyle{plain} 
\theoremstyle{plain} \newtheorem{df}[thm]{Definition}
\theoremstyle{plain} 

\makeatletter
  
  \@addtoreset{equation}{section}
\makeatother

\def\aa{{\alpha}}
\newcommand{\BZ}{{\Bbb Z}}

\begin{document}

\title{A Determinant Formula for a Class of Rational Solutions of Painlev\'e V Equation}
\author{Tetsu Masuda\dag, Yasuhiro Ohta\ddag~and Kenji Kajiwara\dag \\
\dag Department of Electrical Engineering, Doshisha University, \\
Kyotanabe, Kyoto, 610-0321, Japan \\
\ddag Department of Applied Mathematics, Faculty of Engineering, \\
Hiroshima University, \\
1-4-1 Kagamiyama, Higashi-Hiroshima 739-8527, Japan}
\date{}

\maketitle

\begin{abstract}
We give an explicit determinant formula for a class of rational solutions of
the Painlev\'e V equation in terms of the universal characters. 
\end{abstract}

\section{Introduction and Main Result}
It is known that six Painlev\'e equations are in general irreducible,
namely, their solutions cannot be expressed by ``classical functions''
in the sense of Umemura~\cite{Um:100}. However, it is also known that
they admit classical solutions for special values of parameters except
for P$_{\rm I}$. Much effort have been made for the investigation of
classical solutions. As a result, it has been recognized that there are
two classes of classical solutions. One is transcendental classical
solutions expressible in terms of functions of hypergeometric
type. Another one is algebraic or rational solutions. It is also known
that the Painlev\'e equations (except for P$_{\rm I}$) admit action of
the affine Weyl groups as groups of the B\"acklund transformations.  It
is remarkable that such classical solutions are located on special
places from a point of view of symmetry in the parameter
spaces~\cite{O1,O2,O3,O4}. For example, P$_{\rm II}$, P$_{\rm III}$ and
P$_{\rm IV}$, whose symmetry is described by the affine Weyl group of
type $A_1^{(1)}$, $A_1^{(1)}\times A_1^{(1)}$ and $A_2^{(1)}$,
respectively, admit transcendental classical solutions on the reflection
hyperplanes, and rational solutions on the barycenters of Weyl chambers
of the corresponding affine Weyl group.

Umemura et al have investigated the class of solutions on the
barycenters of Weyl chambers and found that (1) these solutions are
expressed by some characteristic polynomials generated by the Toda type
bilinear equations, (2) the coefficients of such polynomials admit
mysterious combinatorial properties~\cite{Um1,Um2,Tane}.  These special
polynomials are sometimes referred as {\it Yablonskii-Vorob'ev
polynomials} for P$_{\rm II}$~\cite{YV}, {\it Okamoto polynomials} for P$_{\rm
IV}$~\cite{O3}, {\it Umemura polynomials} for P$_{\rm III}$, P$_{\rm V}$ and
P$_{\rm VI}$.

One important aspect among such polynomials is that they are expressed
as special cases of the Schur functions.  As is well known, the Schur
functions are characters of the irreducible polynomial representations
of $GL(n)$ and arise as $\tau$-functions of the KP hierarchy~\cite{JM}.
For example, it is known that the special polynomials for P$_{\rm II}$
and P$_{\rm III}$ are expressible by 2-reduced Schur functions, and
those for P$_{\rm IV}$ by 3-reduced Schur
functions~\cite{p2:rat,p3:rat,p4:rat,NY:P4}.

In this paper, we consider P$_{\rm V}$,
\begin{equation}
\frac{d^2y}{dt^2}=
\left(\frac{1}{2y}+\frac{1}{y-1}\right)\left(\frac{dy}{dt}\right)^2-\frac{1}{t}\frac{dy}{dt}
+\frac{(y-1)^2}{2t^2}\left(\kappa_{\infty}^2 y-\frac{\kappa_0^2}{y}\right)
-(\theta+1)\frac{y}{t}-\frac{y(y+1)}{2(y-1)},  \label{P5}
\end{equation}
with parameters $\kappa_{\infty}$, $\kappa_0$ and $\theta$, whose
symmetry is described by the affine Weyl group $W(A_3^{(1)})$. The aim
of this paper is to investigate a class of rational solutions on the
barycenters of Weyl chambers and to present an explicit formula for
them.

By the analogy from the known cases, it is naively expected that they
are expressed in terms of 4-reduced Schur functions. However, our
formula is expressed by a generalization of Schur functions, which is
called the {\it universal characters} and defined as follows~\cite{Koike}.
\begin{df}
Let $p_k=p_k(t^{(1)})$ and $q_k=q_k(t^{(2)})$ , $k\in\BZ$, be two
families of polynomials defined by
\begin{equation}
 \left.
 \begin{array}{@{\,}ll}
  \displaystyle 
   \sum_{k=0}^{\infty}p_k\eta^k=\exp\left(\sum_{j=1}^{\infty}t_j^{(1)}\eta^j\right), 
   \quad p_k=0\ \mbox{for}\ k<0, \\
  \displaystyle 
   \sum_{k=0}^{\infty}q_k\eta^k=\exp\left(\sum_{j=1}^{\infty}t_j^{(2)}\eta^j\right), 
   \quad q_k=0\ \mbox{for}\ k<0,
 \end{array}
 \right.    \label{gf}
\end{equation}
where $t^{(1)}=(t_1^{(1)},t_2^{(1)},\cdots)$ and 
$t^{(2)}=(t_1^{(2)},t_2^{(2)},\cdots)$ are the sets of infinite numbers of variables. 
For any partitions
$\lambda^{(1)}=(\lambda^{(1)}_1,\lambda^{(1)}_2,\cdots,\lambda^{(1)}_n)$
and
$\lambda^{(2)}=(\lambda^{(2)}_1,\lambda^{(2)}_2,\cdots,\lambda^{(2)}_m)$,
the universal character
 $S_{\lambda^{(1)},\lambda^{(2)}}(t^{(1)},t^{(2)})$ 
is defined as
\begin{equation}
S_{\lambda^{(1)},\lambda^{(2)}}(t^{(1)},t^{(2)})
=\det{}^t
\left(q_{\lambda^{(2)}_m}^-,
      q_{\lambda^{(2)}_{m-1}+1}^-,\cdots,
      q_{\lambda^{(2)}_1+m-1}^-,
      p_{\lambda^{(1)}_1-m}^+,
      p_{\lambda^{(1)}_2-m-1}^+,\cdots,
      p_{\lambda^{(1)}_n-m-n+1}^+
\right),     \label{univ:char}
\end{equation}
where 
\begin{equation}
p_j^+={}^t\left(p_j,p_{j+1},\cdots,p_{j+m+n-1}\right), \quad 
q_j^-={}^t\left(q_j,q_{j-1},\cdots,q_{j-m-n+1}\right). 
\end{equation}
\end{df}
\par\medskip
Our main result is stated as follows. 
\begin{thm}\label{main}
For $m,n \in \BZ_{\ge 0}$, we define a family of polynomials 
$S_{m,n}=S_{m,n}(t,s)$ by specializing 
$S_{\lambda^{(1)},\lambda^{(2)}}(t^{(1)},t^{(2)})$ as 
\begin{equation}
\lambda^{(1)}=(n,n-1,\cdots,2,1), \quad \lambda^{(2)}=(m,m-1,\cdots,2,1),
\end{equation}
\begin{equation}
t_j^{(1)}=-\frac{t}{2}+\frac{2s-m+n}{j}, \quad t_j^{(2)}=\frac{t}{2}+\frac{2s-m+n}{j},
\end{equation}
where $s$ is a parameter. For $m,n \in \BZ_{<0}$, we define $S_{m,n}$ through 
\begin{equation}
 \left.
 \begin{array}{@{\,}ll}
  \displaystyle 
   S_{m,n}(t,s)=(-1)^{m(m+1)/2}S_{-m-1,n}(t,s-m-1/2),  \\
  \displaystyle 
   S_{m,n}(t,s)=(-1)^{n(n+1)/2}S_{m,-n-1}(t,s-n-1/2).
 \end{array}
 \right.    \label{neg:S}
\end{equation}
Then, 
\begin{equation}
y=-\frac{S_{m,n-1}(t,s)S_{m-1,n}(t,s)}{S_{m-1,n}(t,s-1)S_{m,n-1}(t,s+1)}, \label{III}
\end{equation}
gives the rational solutions of P$_{\rm V}$ (\ref{P5}) with the parameters 
\begin{equation}
\kappa_{\infty}=s, \quad \kappa_0=s-m+n, \quad \theta=m+n-1, \label{para-III:1}
\end{equation}
and 
\begin{equation}
\kappa_{\infty}=-s, \quad \kappa_0=s-m+n, \quad \theta=m+n-1. \label{para-III:2}
\end{equation}
Similarly, 
\begin{equation}
y=\frac{2n+1}{2m+1}
  \frac{S_{m,n-1}(t,s+1/2)S_{m,n+1}(t,s-1/2)}{S_{m-1,n}(t,s-1/2)S_{m+1,n}(t,s+1/2)},
\end{equation}
gives the rational solutions of P$_{\rm V}$ (\ref{P5}) with the parameters 
\begin{equation}
\kappa_{\infty}=m+1/2, \quad \kappa_0=n+1/2, \quad \theta=2s-m-n-1.
\end{equation}
\end{thm}
\par\medskip
This result covers all the rational solutions obtained by applying the B\"acklund
transformations on the particular solution of P$_{\rm
V}$ (\ref{P5}),
\begin{equation}
y=-1, \quad \kappa_{\infty}=s, \quad \kappa_0=s, \quad \theta=-1.
\end{equation}
\noindent\textbf{Remark.} In ref.~\cite{KLM}, Kitaev et al gave a
complete classification of rational solutions for P$_{\rm V}$. 
Our result covers all the rational solutions of the cases (III) and (IV) 
in their classification. 
The first half corresponds to the case (III) and the other does to (IV). 
Also, Noumi and Yamada presented a determinant formula for a class of 
rational solutions of P$_{\rm V}$ in
terms of 2-reduced Schur functions~\cite{NY:P5}. Our result includes
their formula as a special case, which is explained in Appendix
\ref{ume}.
\par\medskip

This paper is organized as follows.  In Section \ref{sym}, we give a
brief review for the theory of symmetric form of P$_{\rm
V}$~\cite{NY:P4,NY1,NY2}, which provides us with a clear description of
symmetry structure and $\tau$-functions for P$_{\rm V}$.  In Section
\ref{rat}, we construct the rational solutions of P$_{\rm V}$ by using
the theory of symmetric form.  Proof of our result is given in Section
\ref{proof}.  We mention on the relationship between our result and
Yamada's general determinant formula\cite{Ya} of Jacobi-Trudi type
in Section \ref{conc}.

\section{Symmetric Form of Painlev\'e V Equation \label{sym}}
By using the theory of symmetric form for P$_{\rm V}$, it is possible to
describe the structure of B\"acklund transformations in a unified manner
and to construct particular solutions systematically.  In this section, we
summarize the symmetric form of P$_{\rm V}$ following refs.~\cite{NY1,NY2},
and derive bilinear equations satisfied by $\tau$-functions.

\subsection{Symmetric Form of P$_{\rm V}$}
P$_{\rm V}$ (\ref{P5}) is equivalent to the Hamilton system~\cite{Watanabe}
\begin{equation}
q'= \frac{\partial H}{\partial p}, \quad p'=-\frac{\partial H}{\partial q}, \quad '=t\frac{d}{dt},
\label{HS}
\end{equation}
with the Hamiltonian 
\begin{equation}
H=p(p+t)q(q-1)+\aa_2 qt-\aa_3 pq-\aa_1 p(q-1). \label{H}
\end{equation}
In fact, putting 
\begin{equation}
\kappa_{\infty}=\aa_1, \quad \kappa_0=\aa_3, \quad \theta=\aa_2-\aa_0-1, 
\end{equation}
with 
\begin{equation}
\aa_0=1-\aa_1-\aa_2-\aa_3, \label{norm:para}
\end{equation}
we see that equation for $y=1-1/q$ is nothing but P$_{\rm V}$ (\ref{P5}). 
Setting 
\begin{equation}
f_0=\frac{1}{\sqrt{t}}(t+p), \quad f_1=\sqrt{t}q, \quad 
f_2=-\frac{1}{\sqrt{t}}p, \quad f_3=\sqrt{t}(1-q), 
\end{equation}
we obtain the symmetric form of P$_{\rm V}$
\begin{equation}
\left.
 \begin{array}{@{\,}ll}
 \displaystyle 
  f_0'=f_0 f_2(f_1-f_3)+\left(\frac{1}{2}-\aa_2\right)f_0+\aa_0 f_2, \\
 \displaystyle 
  f_1'=f_1 f_3(f_2-f_0)+\left(\frac{1}{2}-\aa_3\right)f_1+\aa_1 f_3, \\
 \displaystyle 
  f_2'=f_2 f_0(f_3-f_1)+\left(\frac{1}{2}-\aa_0\right)f_2+\aa_2 f_0, \\
 \displaystyle 
  f_3'=f_3 f_1(f_0-f_2)+\left(\frac{1}{2}-\aa_1\right)f_3+\aa_3 f_1. 
 \end{array}
\right.     \label{sym:A3}
\end{equation}
We note that the original dependent variable $y$ of P$_{\rm V}$ is
expressed as
\begin{equation}
y=-\frac{f_3}{f_1}. \label{y-f}
\end{equation}
In terms of variables $f_i$ and $\aa_i$, the B\"acklund transformations of
P$_{\rm V}$ are described by a simple form,
\begin{equation}
 \begin{array}{lll}
 \smallskip
 {\displaystyle s_i(\aa_i)=-\aa_i,} & {\displaystyle s_i(\aa_j)=\aa_j+\aa_i~(j=i \pm 1),} & 
 {\displaystyle s_i(\aa_j)=\aa_j~(j \ne i,i \pm 1),} \\ 
 \smallskip
 {\displaystyle s_i(f_i)=f_i,} & {\displaystyle s_i(f_j)=f_j \pm \frac{\aa_i}{f_i}~(j=i \pm 1),} &
 {\displaystyle s_i(f_j)=f_j ~(j \ne i,i \pm 1),} \\
 {\displaystyle \pi(\aa_j)=\aa_{j+1},} & \pi(f_j)=f_{j+1}, & 
 \end{array}
\label{BT:f}
\end{equation}
where the subscripts $i=0,1,2,3$ are understood as elements of
$\BZ/4\BZ$.  These transformations commute with derivation ${}'$
and satisfy the fundamental relations
\begin{equation}
 \left.
  \begin{array}{c}
  \displaystyle 
   s_i^2=1, \quad s_i s_j =s_j s_i~(j \ne i,i \pm 1), \quad s_i s_j s_i=s_j s_i s_j~(j=i \pm 1), \\
  \displaystyle 
   \pi^4=1, \quad \pi s_j=s_{j+1}\pi, 
 \end{array}
\right.     \label{fun.rel}
\end{equation}
which means that transformations $s_i$ $(i=0,1,2,3)$ generate the affine
Weyl group $W(A^{(1)}_3)$, and $s_i$ and $\pi$ generate its
extension including the Dynkin diagram automorphisms.

\subsection{$\tau$-Functions and Bilinear Equations}
In order to obtain simpler transformation properties, we add a correction term
which depends only on $t$ to the Hamiltonian (\ref{H}).  The corrected
Hamiltonian $h_0$ is introduced as
\begin{equation}
 \left.
  \begin{array}{@{\,}ll}
  \displaystyle 
   h_0=f_0 f_1 f_2 f_3
      +\frac{\aa_1+2\aa_2-\aa_3}{4}f_0 f_1
      +\frac{\aa_1+2\aa_2+3\aa_3}{4}f_1 f_2 \\
  \displaystyle 
   \hskip64pt
      -\frac{3\aa_1+2\aa_2+\aa_3}{4}f_2 f_3
      +\frac{ \aa_1-2\aa_2-\aa_3}{4}f_3 f_0 + \frac{(\aa_1+\aa_3)^2}{4},
 \end{array}
\right.
\end{equation}
and we put $h_j=\pi^j(h_0)$. 
Then, we have 
\begin{equation}
s_i(h_j)=h_j~(i \ne j), \quad s_i(h_i)=h_i+\sqrt{t}~\frac{\aa_i}{f_i}, 
\quad \pi(h_i)=h_{i+1}. 
\end{equation}
We also introduce $\tau$-functions $\tau_i$($i=0,1,2,3$) by
\begin{equation}
h_i=\frac{\tau_i'}{\tau_i}. 
\end{equation}
Then, defining the action of $s_i$($i=0,1,2,3$) and $\pi$ on the $\tau$-functions by
\begin{equation}
s_i(\tau_j)=\tau_j~~(i\neq j), \quad s_i(\tau_i)=f_i\frac{\tau_{i-1}\tau_{i+1}}{\tau_i}, \quad 
\pi(\tau_i)=\tau_{i+1},  \label{BT:tau}
\end{equation}
we see that the fundamental relations (\ref{fun.rel}) are preserved,
which implies 
that the B\"acklund transformations can be lifted to the level of
$\tau$-functions. It should be remarked that we have from (\ref{y-f}) and (\ref{BT:tau}) 
\begin{equation}
y=-\frac{\tau_3 s_3(\tau_3)}{\tau_1 s_1(\tau_1)}. \label{y:tau}
\end{equation}

By (\ref{BT:tau}), the B\"acklund transformations (\ref{BT:f}) are lead to
a set of bilinear equations for $\tau$-functions,
\begin{equation}
 \left.
 \begin{array}{@{\,}ll}
  \tau_0 s_0s_1(\tau_1)=s_0(\tau_0)s_1(\tau_1)+\aa_0 \tau_2 \tau_3, \\
  \tau_1 s_1s_0(\tau_0)=s_0(\tau_0)s_1(\tau_1)-\aa_1 \tau_2 \tau_3, \\
  \tau_1 s_1s_2(\tau_2)=s_1(\tau_1)s_2(\tau_2)+\aa_1 \tau_3 \tau_0, \\
  \tau_2 s_2s_1(\tau_1)=s_1(\tau_1)s_2(\tau_2)-\aa_2 \tau_3 \tau_0, \\
  \tau_2 s_2s_3(\tau_3)=s_2(\tau_2)s_3(\tau_3)+\aa_2 \tau_0 \tau_1, \\
  \tau_3 s_3s_2(\tau_2)=s_2(\tau_2)s_3(\tau_3)-\aa_3 \tau_0 \tau_1, \\
  \tau_3 s_3s_0(\tau_0)=s_3(\tau_3)s_0(\tau_0)+\aa_3 \tau_1 \tau_2, \\
  \tau_0 s_0s_3(\tau_3)=s_3(\tau_3)s_0(\tau_0)-\aa_0 \tau_1 \tau_2.
 \end{array}
 \right.    \label{bi:bt}
\end{equation}

Let us define the translation operators $T_i~(i=0,1,2,3)$ by 
\begin{equation}
T_1=\pi s_3s_2s_1, \quad T_2=s_1\pi s_3s_2, \quad T_3=s_2s_1\pi s_3, \quad T_0=s_3s_2s_1\pi, 
\label{def:T}
\end{equation}
which commute with each other and satisfy $T_1T_2T_3T_0=1$. 
These operators act on parameters $\aa_i$ as 
\begin{equation}
T_i(\aa_{i-1})=\aa_{i-1}+1, \quad T_i(\aa_i)=\aa_i-1, \quad T_i(\aa_j)=\aa_j~(j \ne i-1,i),
\end{equation}
and generate the weight lattice of $A_3^{(1)}$.  In terms of $T_i$,
$\tau$-functions in (\ref{bi:bt}) are expressed as
\begin{equation}
  \begin{array}{lll}
\smallskip
   {\displaystyle \tau_1=T_1(\tau_0),}&{\displaystyle \tau_2=T_1T_2(\tau_0)}, &
{\displaystyle \tau_3=T_0^{-1}(\tau_0),} \\
\smallskip
   {\displaystyle s_0(\tau_0)=T_0^{-1}T_1(\tau_0),}&{\displaystyle s_1(\tau_1)=T_2(\tau_0),}&
  {\displaystyle s_2(\tau_2)=T_1T_3(\tau_0),}\\
\smallskip
{\displaystyle s_3(\tau_3)=T_3^{-1}(\tau_0), } &{\displaystyle s_0s_1(\tau_1)=T_1T_2T_0^{-1}(\tau_0),}
&{\displaystyle s_1s_0(\tau_0)=T_2T_0^{-1}(\tau_0),}\\
\smallskip
{\displaystyle s_1s_2(\tau_2)=T_2T_3(\tau_0),}&{\displaystyle s_2s_1(\tau_1)=T_3(\tau_0), }
&{\displaystyle s_2s_3(\tau_3)=T_2^{-1}(\tau_0),}\\
\smallskip
{\displaystyle s_3s_2(\tau_2)=T_1T_0(\tau_0), } & {\displaystyle s_3s_0(\tau_0)=T_1T_3^{-1}(\tau_0),}
&{\displaystyle s_0s_3(\tau_3)=T_1T_3^{-1}T_0^{-1}(\tau_0)}.
 \end{array}
\label{T:tau}
\end{equation}
Furthermore, we can derive bilinear equations of Toda type. 
\begin{prop}
We have 
\begin{equation}
 \left.
 \begin{array}{@{\,}ll}
  \displaystyle 
   T_1(\tau_0)T_1^{-1}(\tau_0)=
   \frac{1}{\sqrt{t}}\left(\frac{1}{2}D_T^2+\frac{3\aa_1+2\aa_2+\aa_3}{4}t\right)
           \tau_0 \cdot \tau_0, \\
  \displaystyle 
   T_2(\tau_0)T_2^{-1}(\tau_0)=
   \frac{1}{\sqrt{t}}\left(\frac{1}{2}D_T^2-\frac{\aa_1-2\aa_2-\aa_3}{4}t\right)
           \tau_0 \cdot \tau_0, \\
  \displaystyle 
   T_3(\tau_0)T_3^{-1}(\tau_0)=
   \frac{1}{\sqrt{t}}\left(\frac{1}{2}D_T^2-\frac{\aa_1+2\aa_2-\aa_3}{4}t\right)
           \tau_0 \cdot \tau_0, \\
  \displaystyle 
   T_0(\tau_0)T_0^{-1}(\tau_0)=
   \frac{1}{\sqrt{t}}\left(\frac{1}{2}D_T^2-\frac{\aa_1+2\aa_2+3\aa_3}{4}t\right)
           \tau_0 \cdot \tau_0,
 \end{array}
 \right.    \label{Toda}
\end{equation}
where $D_T$ is the
Hirota's differential operator defined by
\begin{equation}
D_T^m f\cdot g = \left.\left(\frac{d}{dT}-\frac{d}{dT'}\right)^m f(T)g(T')\right|_{T=T'},
\end{equation}
and $\displaystyle \frac{d}{dT}=t\frac{d}{dt}$.
\end{prop}
{\it Proof.} Using (\ref{BT:f}),(\ref{BT:tau}) and (\ref{def:T}), we have 
\begin{equation}
T_1(\tau_0)T_1^{-1}(\tau_0)=[f_1f_2f_3+(\aa_1+\aa_2)f_1+\aa_1f_3]\tau_0^2. 
\end{equation}
Noticing that 
\begin{equation}
h_0'=\sqrt{t}
     \left(f_1f_2f_3+\frac{\aa_1+2\aa_2-\aa_3}{4}f_1+\frac{\aa_1-2\aa_2-\aa_1}{4}f_3 \right),
\end{equation}
and $f_1+f_3=\sqrt{t}$, we get the first equation in (\ref{Toda}). 
The other equations are obtained in similar way. \hfill\qed

\par\medskip

\noindent\textbf{Remark.} Bilinear equations (\ref{bi:bt}) and (\ref{Toda}) are
overdetermined systems when they are regarded as equations to determine
the $\tau$-functions. However, by construction, the consistency of these
equations is guaranteed.

\subsection{$\tau$-Cocycles}\label{cocy}
For simplicity, we introduce a notation,
\begin{equation}
\tau_{k,l,m,n}=T_1^kT_2^lT_3^mT_0^n(\tau_0).  \label{tau}
\end{equation}
For small $k,l,m,n$, we observe that $\tau_{k,l,m,n}$ are factorized as 
\begin{equation}
\tau_{k,l,m,n}=\phi_{k,l,m,n}~\tau_0 
        \left(\frac{\tau_1}{\tau_0}\right)^k \left(\frac{\tau_2}{\tau_1}\right)^l 
        \left(\frac{\tau_3}{\tau_2}\right)^m \left(\frac{\tau_0}{\tau_3}\right)^n,  \label{tau-phi}
\end{equation}
where $\phi_{k,l,m,n}$ are some functions of $f_i$ and $\aa_i$.
Conversely, if we define $\phi_{k,l,m,n}$ by (\ref{tau-phi}), 
it is shown that $\phi_{k,l,m,n}$'s are polynomials in $f_i$
and $\alpha_i$ for any $k,l,m,n\in\BZ$~\cite{Ya}. 
The functions $\phi_{k,l,m,n}$ are called the $\tau$-cocycles. 

It is easy to see from
(\ref{y:tau}), (\ref{T:tau}), (\ref{tau}) and (\ref{tau-phi}) that we have
\begin{equation}
T_1^kT_2^lT_3^mT_0^n(y)=
-\frac{\phi_{k,l,m,n-1}\phi_{k,l,m-1,n}}{\phi_{k+1,l,m,n}\phi_{k,l+1,m,n}}, 
\quad \mbox{for}\ k,l,m,n \in \BZ.  \label{y:BT}
\end{equation}
Moreover, it follows from (\ref{BT:tau}), (\ref{def:T}), (\ref{tau}) and
(\ref{tau-phi}) that $\phi_{k,l,m,n}$ are determined by the recurrence
relations,
\begin{equation}
 \left.
 \begin{array}{@{\,}ll}
  \displaystyle 
   \phi_{k+1,l,m,n}=
   T_1(\phi_{k,l,m,n})[f_2f_3f_0-(\aa_3+\aa_0)f_0-\aa_0 f_2]^{k-l}
                      (f_3f_0-\aa_0)^{l-m}f_0^{m-n}, \\
  \displaystyle 
   \phi_{k,l+1,m,n}=
   T_2(\phi_{k,l,m,n})[f_3f_0f_1-(\aa_0+\aa_1)f_1-\aa_1 f_3]^{l-m}
                      (f_0f_1-\aa_1)^{m-n}f_1^{1+n-k}, \\
  \displaystyle 
   \phi_{k,l,m+1,n}=
   T_3(\phi_{k,l,m,n})[f_0f_1f_2-(\aa_1+\aa_2)f_2-\aa_2 f_0]^{m-n}
                      (f_1f_2-\aa_2)^{1+n-k}f_2^{k-l}, \\
  \displaystyle 
   \phi_{k,l,m,n+1}=
   T_0(\phi_{k,l,m,n})[f_1f_2f_3-(\aa_2+\aa_3)f_3-\aa_3 f_1]^{1+n-k}
                      (f_2f_3-\aa_3)^{k-l}f_3^{l-m}, 
 \end{array}
 \right.    \label{rec:phi}
\end{equation}
with $\phi_{0,0,0,0}=1$. 

It is possible to write down the bilinear equations for
$\phi_{k,l,m,n}$.  From (\ref{T:tau}),(\ref{tau}) and (\ref{tau-phi}),
bilinear B\"acklund transformations (\ref{bi:bt}) yield to
\begin{equation}
 \left.
 \begin{array}{@{\,}ll}
  \displaystyle 
   \phi_{k,l,m,n}\phi_{k+1,l+1,m,n-1}=
   \phi_{k+1,l,m,n-1}\phi_{k,l+1,m,n}+(\alpha_0-n+k)\phi_{k+1,l+1,m,n}\phi_{k,l,m,n-1}, \\
  \displaystyle 
   \phi_{k+1,l,m,n}\phi_{k,l+1,m,n-1}=
   \phi_{k+1,l,m,n-1}\phi_{k,l+1,m,n}-(\alpha_1-k+l)\phi_{k+1,l+1,m,n}\phi_{k,l,m,n-1}, \\
  \displaystyle 
   \phi_{k+1,l,m,n}\phi_{k,l+1,m+1,n}=
   \phi_{k,l+1,m,n}\phi_{k+1,l,m+1,n}+(\alpha_1-k+l)\phi_{k,l,m,n-1}\phi_{k,l,m,n}, \\
  \displaystyle 
   \phi_{k+1,l+1,m,n}\phi_{k,l,m+1,n}=
   \phi_{k,l+1,m,n}\phi_{k+1,l,m+1,n}-(\alpha_2-l+m)\phi_{k,l,m,n-1}\phi_{k,l,m,n}, \\
  \displaystyle 
   \phi_{k+1,l+1,m,n}\phi_{k,l-1,m,n}=
   \phi_{k+1,l,m+1,n}\phi_{k,l,m-1,n}+(\alpha_2-l+m)\phi_{k,l,m,n}\phi_{k+1,l,m,n}, \\
  \displaystyle 
   \phi_{k,l,m,n-1}\phi_{k+1,l,m,n+1}=
   \phi_{k+1,l,m+1,n}\phi_{k,l,m-1,n}-(\alpha_3-m+n)\phi_{k,l,m,n}\phi_{k+1,l,m,n}, \\
  \displaystyle 
   \phi_{k,l,m,n-1}\phi_{k+1,l,m-1,n}=
   \phi_{k,l,m-1,n}\phi_{k+1,l,m,n-1}+(\alpha_3-m+n)\phi_{k+1,l,m,n}\phi_{k+1,l+1,m,n}, \\
  \displaystyle 
   \phi_{k,l,m,n}\phi_{k+1,l,m-1,n-1}=
   \phi_{k,l,m-1,n}\phi_{k+1,l,m,n-1}-(\alpha_0-n+k)\phi_{k+1,l,m,n}\phi_{k+1,l+1,m,n}.
 \end{array}
 \right.    \label{bi:bt:phi}
\end{equation}
Similarly, the last two equations in (\ref{Toda}) are lead to 
\begin{equation}
 \begin{array}{l}
\smallskip
  \displaystyle 
   \phi_{k,l,m+1,n}\phi_{k,l,m-1,n}\\\smallskip
   \hskip40pt =\displaystyle \frac{1}{\sqrt{t}}
\left(\frac{1}{2}D_T^2+\omega_{k,l,m,n}-\frac{\alpha_1+2\alpha_2-\alpha_3-k-l+3m-n}{4}t\right)
    \phi_{k,l,m,n} \cdot \phi_{k,l,m,n}, \\
\smallskip
  \displaystyle 
   \phi_{k,l,m,n+1}\phi_{k,l,m,n-1}\\
      \hskip40pt =\displaystyle \frac{1}{\sqrt{t}}
\left(\frac{1}{2}D_T^2+\omega_{k,l,m,n}-\frac{\alpha_1+2\alpha_2+3\alpha_3-k-l-m+3n}{4}t\right)
    \phi_{k,l,m,n} \cdot \phi_{k,l,m,n},
 \end{array}
\label{Toda:phi}
\end{equation}
with 
\begin{equation}
\omega_{k,l,m,n}=
(\log \tau_0)''+k\left(\log \frac{\tau_1}{\tau_0}\right)''
               +l\left(\log \frac{\tau_2}{\tau_1}\right)''
               +m\left(\log \frac{\tau_3}{\tau_2}\right)''
               +n\left(\log \frac{\tau_0}{\tau_3}\right)''.
\end{equation}

\section{Construction of Rational Solutions \label{rat}}
In this section, we construct the rational solutions of P$_{\rm V}$ by
using the results in the previous section.

It is obvious that the symmetric form of P$_{\rm V}$ (\ref{sym:A3}) with
(\ref{norm:para}) has a solution,
\begin{equation}
(\aa_0,\aa_1,\aa_2,\aa_3)=\left(\frac{1}{2}-s,s,\frac{1}{2}-s,s \right), \quad 
f_i=\frac{\sqrt{t}}{2}\ \mbox{for}\ i=0,1,2,3,
\label{seed}
\end{equation}
which is on the fixed points with respect to the transformation $\pi^2$ and is
equivalent to the following solution of P$_{\rm V}$,
\begin{equation}
y=-1, \quad \kappa_{\infty}=s, \quad \kappa_0=s, \quad \theta=-1.
\end{equation}
This is the unique rational solution in the fundamental region of the
affine Weyl group $W(A^{(1)}_3)$ in the parameter space, except for the
special cases of transcendental classical solutions~\cite{Watanabe}.
Applying B\"acklund transformations to the seed solution (\ref{seed}),
we obtain the family of rational solutions of P$_{\rm V}$.  Note that we
have
\begin{equation}
T_2^l T_0^l(\aa_0,\aa_1,\aa_2,\aa_3)=
\left(\frac{1}{2}-\tilde{s},\tilde{s},\frac{1}{2}-\tilde{s},\tilde{s} \right), \quad \tilde{s}=s+l,
\quad l \in \BZ, 
\label{seed'}
\end{equation}
under the specialization (\ref{seed}).  Comparing (\ref{seed'}) with
(\ref{seed}), we see that the effect of $T_2$ is absorbed by that of
$T_0^{-1}$ and shift of the parameter $s$.  Then, we do not need to consider
the B\"acklund transformation $T_2$ for constructing the family of
rational solutions of P$_{\rm V}$.  Taking the initial condition
(\ref{seed}) into account, we consider the bilinear B\"acklund
transformations in terms of the $\tau$-cocycles (\ref{bi:bt:phi}).
Denoting $\phi_{0,l,m,n}=\phi_{l,m,n}$ in view of the relation $T_1T_2T_3T_0=1$, it is
easy to derive the following.

\begin{lem}\label{y:phi}
Under the specialization (\ref{seed}), bilinear B\"acklund
transformations (\ref{bi:bt:phi}) are reduced to
\begin{equation}
 \left.
 \begin{array}{@{\,}ll}
  \displaystyle 
   \phi_{0,m,n}\phi_{0,m-1,n-2}=
   \phi_{-1,m-1,n-2}\phi_{1,m,n}+(1/2-s-n)\phi_{0,m-1,n-1}\phi_{0,m,n-1}, \\
  \displaystyle 
   \phi_{-1,m-1,n-1}\phi_{1,m,n-1}=
   \phi_{-1,m-1,n-2}\phi_{1,m,n}-s\phi_{0,m-1,n-1}\phi_{0,m,n-1}, \\
  \displaystyle 
   \phi_{-1,m-1,n-1}\phi_{1,m+1,n}=
   \phi_{1,m,n}\phi_{-1,m,n-1}+s\phi_{0,m,n-1}\phi_{0,m,n}, \\
  \displaystyle 
   \phi_{0,m-1,n-1}\phi_{0,m+1,n}=
   \phi_{1,m,n}\phi_{-1,m,n-1}-(1/2-s+m)\phi_{0,m,n-1}\phi_{0,m,n}, \\
  \displaystyle 
   \phi_{0,m-1,n-1}\phi_{-1,m,n}=
   \phi_{-1,m,n-1}\phi_{0,m-1,n}+(1/2-s+m)\phi_{0,m,n}\phi_{-1,m-1,n-1}, \\
  \displaystyle 
   \phi_{0,m,n-1}\phi_{-1,m-1,n}=
   \phi_{-1,m,n-1}\phi_{0,m-1,n}-(s-m+n)\phi_{0,m,n}\phi_{-1,m-1,n-1}, \\
  \displaystyle 
   \phi_{0,m,n-1}\phi_{-1,m-2,n-1}=
   \phi_{0,m-1,n}\phi_{-1,m-1,n-2}+(s-m+n)\phi_{-1,m-1,n-1}\phi_{0,m-1,n-1}, \\
  \displaystyle 
   \phi_{0,m,n}\phi_{-1,m-2,n-2}=
   \phi_{0,m-1,n}\phi_{-1,m-1,n-2}-(1/2-s-n)\phi_{-1,m-1,n-1}\phi_{0,m-1,n-1}. 
 \end{array}
 \right.    \label{bi:bt:phi'}
\end{equation}
Moreover, from (\ref{y:BT}), the function 
\begin{equation}
y=-\frac{\phi_{0,m,n-1}\phi_{0,m-1,n}}{\phi_{-1,m-1,n-1}\phi_{1,m,n}},
\end{equation}
solves P$_{\rm V}$ (\ref{P5}) with parameters 
\begin{equation}
\kappa_{\infty}=s, \quad \kappa_0=s-m+n, \quad \theta=m+n-1.
\end{equation}
\end{lem}

From the recurrence relations (\ref{rec:phi}), we observe that
$\phi_{l,m,n}$ for small $l,m,n$ are expressed as
\begin{equation}
\phi_{l,m,n}=\left(\frac{\sqrt{t}}{2}\right)^{(m-n-l-1)(m-n-l)/2}U_{l,m,n}, \label{phi-U}
\end{equation}
where $U_{l,m,n}$ are some polynomials in $t$ and $s$. Therefore, we
next rewrite (\ref{bi:bt:phi'}) in terms of $U$. 
The polynomials $U_{l,m,n}$ have symmetry described by the following lemma.
\begin{lem}\label{sym:U}
The polynomials $U_{l,m,n}$ defined by (\ref{phi-U}) satisfy
\begin{equation}
U_{1,m,n}(t,s)=U_{0,m,n-1}(t,s+1), \quad U_{-1,m,n}(t,s)=U_{0,m,n+1}(t,s-1). \label{Upm}
\end{equation} 
\end{lem}
{\it Proof.} Lemma \ref{sym:U} is proved by considering the Toda type
equation for $\tau$-cocycles. Under the specialization (\ref{seed}),
the Hamiltonians and $\tau$-functions are calculated as 
\begin{equation}
h_0=h_2=\frac{t^2}{16}+s^2, \quad h_1=h_3=\frac{t^2}{16}+\left(\frac{1}{2}-s\right)^2,
\end{equation}
and 
\begin{equation}
\tau_0=\tau_2=t^{s^2}\exp\left(\frac{t^2}{32}\right), \quad 
\tau_1=\tau_3=t^{(1/2-s)^2}\exp\left(\frac{t^2}{32}\right),
\end{equation}
up to the multiplication by some constants, respectively. 
Then, Toda type bilinear equations (\ref{Toda:phi}) yield to 
\begin{equation}
 \left.
 \begin{array}{@{\,}ll}
  \displaystyle 
   \phi_{l,m+1,n}\phi_{l,m-1,n}=\frac{1}{\sqrt{t}}
    \left(\frac{1}{2}D_T^2+\frac{t^2}{8}-\frac{-2s+1-l+3m-n}{4}t\right)
    \phi_{l,m,n} \cdot \phi_{l,m,n}, \\
  \displaystyle 
   \phi_{l,m,n+1}\phi_{l,m,n-1}=\frac{1}{\sqrt{t}}
    \left(\frac{1}{2}D_T^2+\frac{t^2}{8}-\frac{2s+1-l-m+3n}{4}t\right)
    \phi_{l,m,n} \cdot \phi_{l,m,n}.
 \end{array}
 \right.    \label{Toda:phi'}
\end{equation}
Substituting (\ref{phi-U}) into (\ref{Toda:phi'}), 
we obtain Toda type bilinear equations to be satisfied by $U_{m,n}=U_{0,m,n}(t,s)$
\begin{equation}
 \left.
 \begin{array}{@{\,}ll}
  \displaystyle 
   U_{m+1,n}U_{m-1,n}= \\
  \displaystyle 
   \hskip20pt
   2t\left[\left(\frac{d^2U_{m,n}}{dt^2}\right)U_{m,n}-\left(\frac{dU_{m,n}}{dt}\right)^2\right]
   +2\frac{dU_{m,n}}{dt}U_{m,n}
   +\left(\frac{t}{4}-\frac{-2s+1+3m-n}{2}\right)U_{m,n}^2, \\
  \displaystyle 
   U_{m,n+1}U_{m,n-1}= \\
  \displaystyle 
   \hskip20pt
   2t\left[\left(\frac{d^2U_{m,n}}{dt^2}\right)U_{m,n}-\left(\frac{dU_{m,n}}{dt}\right)^2\right]
   +2\frac{dU_{m,n}}{dt}U_{m,n}
   +\left(\frac{t}{4}-\frac{2s+1-m+3n}{2}\right)U_{m,n}^2,
 \end{array}
 \right.    \label{Toda:U}
\end{equation}
with initial conditions 
\begin{equation}
U_{-1,-1}=U_{-1,0}=U_{0,-1}=U_{0,0}=1. \label{init}
\end{equation}
The functions $U_{m,n}=U_{m,n}(t,s)$ are uniquely determined by Toda equations
(\ref{Toda:U}) from the initial conditions (\ref{init}) for any $m,n \in
\BZ$.  Moreover, we see that $U_{\pm 1,m,n}(t,s)$ satisfy the
same Toda equations as $U_{0,m,n \mp 1}(t,s \pm 1)$, respectively, by
the similar calculation.  Since the
initial conditions for $U_{1,m,n}$ and $U_{-1,m,n}$ are given by
\begin{equation}
 \left.
 \begin{array}{c}
  U_{1,-1,0}=U_{1,0,0}=U_{1,-1,1}=U_{1,0,1}=1,         \\
  U_{-1,-1,-2}=U_{-1,-1,-1}=U_{-1,0,-2}=U_{-1,0,-1}=1,
 \end{array}
 \right.
\end{equation}
the lemma is proved.\hfill\qed

From Lemma \ref{sym:U}, bilinear B\"acklund
transformations (\ref{bi:bt:phi'}) are rewritten in terms of $U$.
\begin{prop}\label{y:U}
Let $U_{m,n}=U_{m,n}(t,s)~(m,n \in \BZ)$ be polynomials which
satisfy the bilinear equations,
\begin{equation}
 \left.
 \begin{array}{@{\,}ll}
  4U_{m,n+1}U_{m-1,n-1}=tU^-_{m-1,n}U^+_{m,n}-2(2s+2n+1)U_{m-1,n}U_{m,n}, \\
  4U^-_{m-1,n+1}U^+_{m,n-1}=tU^-_{m-1,n}U^+_{m,n}-4sU_{m-1,n}U_{m,n}, \\
  4U^-_{m-1,n}U^+_{m+1,n-1}=tU^+_{m,n-1}U^-_{m,n}+4s U_{m,n-1}U_{m,n}, \\
  4U_{m-1,n-1}U_{m+1,n}=tU^+_{m,n-1}U^-_{m,n}+2(2s-2m-1)U_{m,n-1}U_{m,n},  \\
  4U_{m-1,n-1}U^-_{m,n+1}=tU^-_{m,n}U_{m-1,n}-2(2s-2m-1)U_{m,n}U^-_{m-1,n}, \\
  4U_{m,n-1}U^-_{m-1,n+1}=tU^-_{m,n}U_{m-1,n}-4(s-m+n)U_{m,n}U^-_{m-1,n}, \\
  4U_{m+1,n-1}U^-_{m-1,n}=tU_{m,n}U^-_{m,n-1}+4(s-m+n-1)U^-_{m,n}U_{m,n-1}, \\
  4U_{m+1,n}U^-_{m-1,n-1}=tU_{m,n}U^-_{m,n-1}+2(2s+2n-1)U^-_{m,n}U_{m,n-1},
 \end{array}
 \right.    \label{bi:bt:U}
\end{equation}
with 
\begin{equation}
U_{-1,-1}=U_{-1,0}=U_{0,-1}=U_{0,0}=1,
\end{equation}
where we denote $U^{\pm}_{m,n}=U_{m,n}(t,s \pm 1)$. 
Then, 
\begin{equation}
y=-\frac{U_{m,n-1}(t,s)U_{m-1,n}(t,s)}{U_{m-1,n}(t,s-1)U_{m,n-1}(t,s+1)},
\end{equation}
gives the rational solutions of P$_{\rm V}$ (\ref{P5}) with parameters 
\begin{equation}
\kappa_{\infty}=s, \quad \kappa_0=s-m+n, \quad \theta=m+n-1. \label{para}
\end{equation}
\end{prop}

\section{Proof of Theorem \ref{main}\label{proof}}
In this section, we give the proof for Theorem \ref{main}.
\begin{df}
Let $p_k^{(r)}=p_k^{(r)}(x)$ and $q_k^{(r)}=q_k^{(r)}(x)$ be 
polynomials defined by
\begin{equation}
\sum_{k=0}^{\infty}p_k^{(r)}\eta^k=(1-\eta)^{-r}\exp\left(-\frac{x\eta}{1-\eta}\right), 
\quad p_k^{(r)}=0\ \mbox{for}\ k<0,
\end{equation}
\begin{equation}
 q_k^{(r)}(x)=p_k^{(r)}(-x),
\end{equation}
respectively.  For $m,n \in \BZ_{\ge 0}$,
we define a family of polynomials $R_{m,n}^{(r)}=R_{m,n}^{(r)}(x)$ by
\begin{equation}
R_{m,n}^{(r)}(x)=
 \left|
  \begin{array}{cccccccc}
   q_1^{(r)}        & q_0^{(r)}        & \cdots           & q_{-m+2}^{(r)}   &
   q_{-m+1}^{(r)}   & \cdots           & q_{-m-n+3}^{(r)} & q_{-m-n+2}^{(r)} \\
   q_3^{(r)}        & q_2^{(r)}        & \cdots           & q_{-m+4}^{(r)}   &
   q_{-m+3}^{(r)}   & \cdots           & q_{-m-n+5}^{(r)} & q_{-m-n+4}^{(r)} \\
   \vdots           & \vdots           & \ddots           & \vdots           &
   \vdots           & \ddots           & \vdots           & \vdots           \\
   q_{2m-1}^{(r)}   & q_{2m-2}^{(r)}   & \cdots           & q_m^{(r)}        &
   q_{m-1}^{(r)}    & \cdots           & q_{m-n+1}^{(r)}  & q_{m-n}^{(r)}    \\
   p_{n-m}^{(r)}    & p_{n-m+1}^{(r)}  & \cdots           & p_{n-1}^{(r)}    &
   p_n^{(r)}        & \cdots           & p_{2n-2}^{(r)}   & p_{2n-1}^{(r)}   \\
   \vdots           & \vdots           & \ddots           & \vdots           &
   \vdots           & \ddots           & \vdots           & \vdots           \\
   p_{-n-m+4}^{(r)} & p_{-n-m+5}^{(r)} & \cdots           & p_{-n+3}^{(r)}   &
   p_{-n+4}^{(r)}   & \cdots           & p_2^{(r)}        & p_3^{(r)}        \\
   p_{-n-m+2}^{(r)} & p_{-n-m+3}^{(r)} & \cdots           & p_{-n+1}^{(r)}   &
   p_{-n+2}^{(r)}   & \cdots           & p_0^{(r)}        & p_1^{(r)}
  \end{array}
 \right|.    \label{2-2:Sch}
\end{equation}
For $m,n \in \BZ_{<0}$, we define $R_{m,n}^{(r)}$ through 
\begin{equation}
R_{m,n}^{(r)}=(-1)^{m(m+1)/2}R_{-m-1,n}^{(r)}, \quad 
R_{m,n}^{(r)}=(-1)^{n(n+1)/2}R_{m,-n-1}^{(r)}. \label{neg:R}
\end{equation}
\end{df}
\noindent\textbf{Remark.}
The polynomials $p_k$ and $q_k$ $(k\geq 0)$ are essentially the Laguerre polynomials,
namely, $p_k^{(r)}(x)=L_k^{(r-1)}(x)$.  Moreover, $R_{m,n}^{(r)}$ is
related to $S_{m,n}$ in Theorem \ref{main} as
\begin{equation}
R_{m,n}^{(r)}(x)=S_{m,n}(t,s),\quad  x=\frac{t}{2}, \quad r=2s-m+n. \label{xtrs}
\end{equation}
\par\medskip
\begin{prop}\label{bi:rel}
For $m,n \in \BZ$, $R_{m,n}^{(r)}$ satisfy the following bilinear equations.
\begin{equation}
 \left.
 \begin{array}{@{\,}ll}
\smallskip
 -(2n+1)R_{m,n+1}^{(r+1)}R_{m-1,n-1}^{(r)}
  =xR_{m-1,n}^{(r-1)}R_{m,n}^{(r+2)}-(r+m+n+1)R_{m-1,n}^{(r+1)}R_{m,n}^{(r)}, \\
\smallskip
 -(2n+1)R_{m-1,n+1}^{(r)}R_{m,n-1}^{(r+1)}
  =xR_{m-1,n}^{(r-1)}R_{m,n}^{(r+2)}-(r+m-n)R_{m-1,n}^{(r+1)}R_{m,n}^{(r)}, \\
\smallskip
  (2m+1)R_{m-1,n}^{(r-1)}R_{m+1,n-1}^{(r)}
  =xR_{m,n-1}^{(r+1)}R_{m,n}^{(r-2)}+(r+m-n)R_{m,n-1}^{(r-1)}R_{m,n}^{(r)}, \\
\smallskip
  (2m+1)R_{m-1,n-1}^{(r)}R_{m+1,n}^{(r-1)}
  =xR_{m,n-1}^{(r+1)}R_{m,n}^{(r-2)}+(r-m-n-1)R_{m,n-1}^{(r-1)}R_{m,n}^{(r)}, \\
\smallskip
 -(2n+1)R_{m-1,n-1}^{(r)}R_{m,n+1}^{(r-1)}
  =xR_{m,n}^{(r-2)}R_{m-1,n}^{(r+1)}-(r-m-n-1)R_{m,n}^{(r)}R_{m-1,n}^{(r-1)}, \\
\smallskip
 -(2n+1)R_{m,n-1}^{(r-1)}R_{m-1,n+1}^{(r)}
  =xR_{m,n}^{(r-2)}R_{m-1,n}^{(r+1)}-(r-m+n)R_{m,n}^{(r)}R_{m-1,n}^{(r-1)}, \\
\smallskip
  (2m+1)R_{m+1,n-1}^{(r-2)}R_{m-1,n}^{(r-1)}
  =xR_{m,n}^{(r)}R_{m,n-1}^{(r-3)}+(r-m+n-2)R_{m,n}^{(r-2)}R_{m,n-1}^{(r-1)}, \\
\smallskip
  (2m+1)R_{m+1,n}^{(r-1)}R_{m-1,n-1}^{(r-2)}
  =xR_{m,n}^{(r)}R_{m,n-1}^{(r-3)}+(r+m+n-1)R_{m,n}^{(r-2)}R_{m,n-1}^{(r-1)}.
 \end{array}
 \right.    \label{bi:bt:R}
\end{equation}
\end{prop}
Comparing (\ref{bi:bt:R}) with (\ref{bi:bt:U}), we obtain
an explicit formula for $U_{m,n}=U_{m,n}(t,s)$ in Proposition \ref{y:U}. 
\begin{prop}\label{U:S}
We have 
\begin{equation}
U_{m,n}(t,s)=c_m d_n S_{m,n}(t,s), \quad m,n \in \BZ, \label{U-S}
\end{equation}
where $c_m$ and $d_n$ are constants determined by 
\begin{equation}
 \left.
 \begin{array}{c}
  \displaystyle 
   c_{m+1}c_{m-1}= \left(m+\frac{1}{2}\right)c_m^2, \quad c_{-1}=c_0=1, \\
  \displaystyle 
   d_{n+1}d_{n-1}=-\left(n+\frac{1}{2}\right)d_n^2, \quad d_{-1}=d_0=1.
 \end{array}
 \right.
\end{equation}
\end{prop}
{\it Proof.}
Putting 
\begin{equation}
U_{m,n}(t,s)=c_m d_n R_{m,n}^{(r)}(x), 
\end{equation}
with 
\begin{equation}
x=\frac{t}{2}, \quad r=2s-m+n,
\end{equation}
we find that the bilinear relations (\ref{bi:bt:R}) become
(\ref{bi:bt:U}).  Taking (\ref{xtrs}) into account, we obtain
Proposition \ref{U:S}.\hfill\qed\\

Applying $s_1$ to the solutions (\ref{III}) with (\ref{para-III:1}), we get the solutions (\ref{III}) with (\ref{para-III:2}). 
Then, the first half of Theorem \ref{main} is a direct consequence of Proposition \ref{y:U} and \ref{U:S}. 
It is easy to find that the latter half of Theorem \ref{main} is obtained by applying $\pi s_1$ to the solutions (\ref{III}) with (\ref{para-III:1}). 
Therefore, now the proof of Theorem \ref{main} is reduced to that of Proposition \ref{bi:rel}.
\par\medskip
It is possible to reduce the number of bilinear equations to be proved in (\ref{bi:bt:R})
by the following symmetry of $R_{m,n}^{(r)}(x)$.
\begin{lem}\label{simple}
We have the relations for $m,n \in \BZ_{\ge 0}$ 
\begin{equation}
R_{n,m}^{(r)}(-x)=R_{m,n}^{(r)}(x), \label{x->-x}
\end{equation}
\begin{equation}
R_{n,m}^{(-r)}(x)=(-1)^{m(m+1)/2+n(n+1)/2}R_{m,n}^{(r)}(x).
\label{r->-r}
\end{equation}
\end{lem}
{\it Proof.}  The first relation (\ref{x->-x}) is easily obtained from
(\ref{2-2:Sch}).  To verify the second relation (\ref{r->-r}), we
introduce polynomials $\bar{q}_k^{(r)}=\bar{q}_k^{(r)}(x)$ by
\begin{equation}
\sum_{k=0}^{\infty}\bar{q}_k^{(r)}\eta^k
=(1+\eta)^r\exp\left(\frac{x\eta}{1+\eta}\right), 
\quad \bar{q}_k^{(r)}=0\ \mbox{for}\ k<0. 
\end{equation}
Comparing the generating function for $q_k$ with that for $\bar{q}_k$, we see that
each $\bar{q}_k^{(r)}(x)$ is a linear combination of
$q_j^{(r)}(x),~j=k,k-2,k-4,\cdots$. Therefore we can express $R_{m,n}^{(r)}$ for $m,n
\in \BZ_{\ge 0}$ in terms of $p_k$ and $\bar{q}_k$ as
\begin{equation}
R_{m,n}^{(r)}(x)=
 \left|
  \begin{array}{cccccccc}
   \bar{q}_1^{(r)}      & \bar{q}_0^{(r)}      & \cdots                 & \bar{q}_{-m+2}^{(r)}   &
   \bar{q}_{-m+1}^{(r)} & \cdots               & \bar{q}_{-m-n+3}^{(r)} & \bar{q}_{-m-n+2}^{(r)} \\
   \bar{q}_3^{(r)}      & \bar{q}_2^{(r)}      & \cdots                 & \bar{q}_{-m+4}^{(r)}   &
   \bar{q}_{-m+3}^{(r)} & \cdots               & \bar{q}_{-m-n+5}^{(r)} & \bar{q}_{-m-n+4}^{(r)} \\
   \vdots               & \vdots               & \ddots                 & \vdots                 &
   \vdots               & \ddots               & \vdots                 & \vdots                 \\
   \bar{q}_{2m-1}^{(r)} & \bar{q}_{2m-2}^{(r)} & \cdots                 & \bar{q}_m^{(r)}        &
   \bar{q}_{m-1}^{(r)}  & \cdots               & \bar{q}_{m-n+1}^{(r)}  & \bar{q}_{m-n}^{(r)}    \\
   p_{n-m}^{(r)}    & p_{n-m+1}^{(r)}  & \cdots           & p_{n-1}^{(r)}    &
   p_n^{(r)}        & \cdots           & p_{2n-2}^{(r)}   & p_{2n-1}^{(r)}   \\
   \vdots           & \vdots           & \ddots           & \vdots           &
   \vdots           & \ddots           & \vdots           & \vdots           \\
   p_{-n-m+4}^{(r)} & p_{-n-m+5}^{(r)} & \cdots           & p_{-n+3}^{(r)}   &
   p_{-n+4}^{(r)}   & \cdots           & p_2^{(r)}        & p_3^{(r)}        \\
   p_{-n-m+2}^{(r)} & p_{-n-m+3}^{(r)} & \cdots           & p_{-n+1}^{(r)}   &
   p_{-n+2}^{(r)}   & \cdots           & p_0^{(r)}        & p_1^{(r)}
  \end{array}
 \right|.
\end{equation}
Noticing that $\bar{q}_k$ and $p_k$ are related as
\begin{equation}
\bar{q}_k^{(r)}(x)=(-1)^k p_k^{(-r)}(x),
\end{equation}
we obtain the relation (\ref{r->-r}). \hfill\qed

From the symmetries of $R_{m,n}^{(r)}(x)$ described by (\ref{neg:R}) and
Lemma \ref{simple}, it is sufficient to prove the first two equations in
(\ref{bi:bt:R}) for $m,n \in \BZ_{\ge 0}$, which are equivalent to
\begin{equation}
R_{m-1,n+1}^{(r)}R_{m,n-1}^{(r+1)}-R_{m,n+1}^{(r+1)}R_{m-1,n-1}^{(r)}
-R_{m-1,n}^{(r+1)}R_{m,n}^{(r)}=0, \label{bt:1}
\end{equation}
\begin{equation}
-(2n+1)R_{m-1,n+1}^{(r)}R_{m,n-1}^{(r+1)}
=xR_{m-1,n}^{(r-1)}R_{m,n}^{(r+2)}-(r+m-n)R_{m-1,n}^{(r+1)}R_{m,n}^{(r)}.
\label{bt:org}
\end{equation}
In the following, we show that these bilinear equations are reduced to
Jacobi's identity of determinants.  Let $D$ be an $(m+n+1) \times
(m+n+1)$ determinant and $\displaystyle D\left[\begin{array}{cccc} i_1 &
i_2 & \cdots & i_k \\ j_1 & j_2 & \cdots & j_k
       \end{array}
 \right]$ the minor which are obtained by deleting the rows with indices
$i_1,\cdots,i_k$ and the columns with indices $j_1,\cdots,j_k$.  Then we
have Jacobi's identity
\begin{equation}
D \cdot D\left[\begin{array}{cc}
           m &  m+1  \\
           1 & m+n+1
	  \end{array}
   \right]=
D\left[\begin{array}{c}
        m \\
        1
       \end{array}
  \right]
D\left[\begin{array}{c}
         m+1  \\
        m+n+1
       \end{array}
 \right]-
D\left[\begin{array}{c}
        m+1 \\
         1
       \end{array}
 \right]
D\left[\begin{array}{c}
          m   \\
        m+n+1
       \end{array}
 \right].         \label{Jacobi}
\end{equation}
We first choose proper determinants as $D$ ($D$ itself should be expressed
in terms of $R_{m,n}^{(r)}$). Secondly, we construct such
formulas that express the minor determinants by $R_{m,n}^{(r)}$.
Then, Jacobi's identity yields bilinear equations for $R_{m,n}^{(r)}$
which are nothing but (\ref{bt:1}) and (\ref{bt:org}).

We have the following lemmas. 
\begin{lem}\label{dif:form:1}
We put 
\begin{equation}
D \equiv
 \left|
  \begin{array}{cccccc}
   -q_1^{(r+1)}       & q_1^{(r)}        & \cdots & q_{-m-n+3}^{(r)} & q_{-m-n+2}^{(r)} \\
   -q_3^{(r+1)}       & q_3^{(r)}        & \cdots & q_{-m-n+5}^{(r)} & q_{-m-n+4}^{(r)} \\
   \vdots             & \vdots           & \ddots & \vdots           & \vdots           \\
   -q_{2m-1}^{(r+1)}  & q_{2m-1}^{(r)}   & \cdots & q_{m-n+1}^{(r)}  & q_{m-n}^{(r)}    \\
   p_{n-m+1}^{(r+1)}  & p_{n-m+2}^{(r)}  & \cdots & p_{2n}^{(r)}     & p_{2n+1}^{(r)}   \\
   \vdots             & \vdots           & \ddots & \vdots           & \vdots           \\
   p_{-n-m+3}^{(r+1)} & p_{-n-m+4}^{(r)} & \cdots & p_2^{(r)}        & p_3^{(r)}        \\
   p_{-n-m+1}^{(r+1)} & p_{-n-m+2}^{(r)} & \cdots & p_0^{(r)}        & p_1^{(r)}
  \end{array}
 \right|\ .
\end{equation}
Then, we have 
\begin{equation}
 \left.
 \begin{array}{@{\,}ll}
  \displaystyle
   D=(-1)^m R_{m,n+1}^{(r+1)}, \quad 
  \displaystyle
   D\left[\begin{array}{c}
             m \\
             1
          \end{array}
    \right]
   =R_{m-1,n+1}^{(r)}, \\
  \displaystyle
   D\left[\begin{array}{c}
            m+1 \\
             1
          \end{array}
    \right]
   =R_{m,n}^{(r)}, \quad 
  \displaystyle
   D\left[\begin{array}{c}
             m \\
           m+n+1
          \end{array}
    \right]
   =(-1)^{m-1}R_{m-1,n}^{(r+1)}, \\
  \displaystyle
   D\left[\begin{array}{c}
            m+1  \\
           m+n+1
          \end{array}
    \right]
   =(-1)^mR_{m,n-1}^{(r+1)}, \quad 
  \displaystyle
   D\left[\begin{array}{cc}
             m &  m+1  \\
             1 & m+n+1
           \end{array}
     \right]
   =R_{m-1,n-1}^{(r)}.
 \end{array}
 \right.
\end{equation}
\end{lem}
\begin{lem}\label{dif:form:2}
We put 
\begin{equation}
D \equiv
 \left|
 \begin{array}{ccccc}
  \tilde{q}_1^{(r-m-n+2)}      & q_1^{(r-m-n+1)}      & q_0^{(r-m-n+2)}      & \cdots
                               & q_{-m-n+2}^{(r)} \\
  \tilde{q}_3^{(r-m-n+2)}      & q_3^{(r-m-n+1)}      & q_2^{(r-m-n+2)}      & \cdots
                               & q_{-m-n+4}^{(r)} \\
  \vdots                       & \vdots               & \vdots               & \ddots & \vdots  \\
  \tilde{q}_{2m-1}^{(r-m-n+2)} & q_{2m-1}^{(r-m-n+1)} & q_{2m-2}^{(r-m-n+2)} & \cdots
                               & q_{m-n}^{(r)}    \\
  (-1)^{m+n}\hat{p}_{2n}^{(r-m-n+2)} & (-1)^{m+n}  p_{2n+1}^{(r-m-n+1)}
  & (-1)^{m+n-1}p_{2n+1}^{(r-m-n+2)} & \cdots & (-1)^1 p_{2n+1}^{(r)}   \\
  \vdots                             & \vdots
  & \vdots                           & \ddots & \vdots                  \\
  (-1)^{m+n}\hat{p}_2^{(r-m-n+2)}    & (-1)^{m+n}p_3^{(r-m-n+1)}
  & (-1)^{m+n-1}p_3^{(r-m-n+2)}      & \cdots & (-1)^1 p_3^{(r)}        \\
  (-1)^{m+n}\hat{p}_0^{(r-m-n+2)}    & (-1)^{m+n}p_1^{(r-m-n+1)}
  & (-1)^{m+n-1}p_1^{(r-m-n+2)}      & \cdots & (-1)^1 p_1^{(r)}
 \end{array}
 \right|, 
\end{equation}
where $\hat{p}_{2k}^{(r)}$ and $\tilde{q}_{2k-1}^{(r)}$ are defined by 
\begin{equation}
\hat{p}_{2k}^{(r)}=\frac{p_{2k}^{(r)}}{2k+1}, \quad 
\tilde{q}_{2k-1}^{(r)}=\frac{q_{2k-1}^{(r)}}{r+2k-2}. 
\end{equation}
Then, we have 
\begin{equation}
 \left.
 \begin{array}{@{\,}ll}
  \displaystyle
   D=\frac{x^{m+n}}{\displaystyle \prod_{j=0}^n(2j+1)\prod_{k=1}^m(r-m-n+2k)}R_{m,n}^{(r+2)}, \\
  \displaystyle
   D\left[\begin{array}{c}
             m \\
             1
          \end{array}
    \right]
   =(-1)^{n+1}R_{m-1,n+1}^{(r)}, 
  \quad 
   D\left[\begin{array}{c}
            m+1 \\
             1
          \end{array}
    \right]
   =(-1)^nR_{m,n}^{(r)}, \\
  \displaystyle
   D\left[\begin{array}{c}
             m   \\
           m+n+1
          \end{array}
    \right]
   =(-1)^{n+1}\frac{x^{m+n-1}}{\displaystyle \prod_{j=0}^n(2j+1)\prod_{k=1}^{m-1}(r-m-n+2k)}
    R_{m-1,n}^{(r+1)}, \\
  \displaystyle
   D\left[\begin{array}{c}
            m+1  \\
           m+n+1
          \end{array}
    \right]
   =(-1)^n\frac{x^{m+n-1}}{\displaystyle \prod_{j=0}^{n-1}(2j+1)\prod_{k=1}^m(r-m-n+2k)}
    R_{m,n-1}^{(r+1)}, \\
  \displaystyle
   D\left[\begin{array}{cc}
             m &  m+1  \\
             1 & m+n+1
           \end{array}
     \right]
   =R_{m-1,n}^{(r-1)}.
 \end{array}
 \right.
\end{equation}
\end{lem}
It is easy to see that the bilinear relations (\ref{bt:1}) and
(\ref{bt:org}) follow immediately from Jacobi's identity (\ref{Jacobi})
by using Lemmas \ref{dif:form:1} and \ref{dif:form:2}, respectively.  We give the
proof of Lemmas \ref{dif:form:1} and \ref{dif:form:2} in Appendix
\ref{PL}.  This completes the proof of Proposition \ref{bi:rel} and thus
our main result Theorem \ref{main}.

\section{Discussion \label{conc}}
As we mentioned in Section \ref{cocy}, the $\tau$- cocycles $\phi_{k,l,m,n}$ are polynomials
in $\aa_i$ and $f_i$ and admit a determinant expression in terms of a
generalized Jacobi-Trudi formula~\cite{Ya}.  When specialized to the
seed solution (\ref{seed}), we obtain a determinant formula for
$\phi_{0,0,m,n}=\phi_{m,n}$, which are given as follows.
\begin{prop}
Let $g_k^{(l)}~(k,l \in \BZ)$ be functions defined by 
\begin{equation}
\left.
 \begin{array}{@{\,}ll}
 \displaystyle 
  g_{2k}^{(2l)}  =    \frac{(-1)^k}{\xi_k}           L_k^{(2s-1)}(t/2), \quad 
  g_{2k+1}^{(2l)}=\frac{\sqrt{t}}{2}\frac{(-1)^k}{\xi_{k+1}^{\dag}}L_k^{(2s^{\dag})}(t/2), \\
 \displaystyle 
  g_{2k}^{(2l+1)}  =    \frac{(-1)^k}{\xi_k^{\dag}}L_k^{(2s^{\dag}-1)}(t/2), \quad 
  g_{2k+1}^{(2l+1)}=\frac{\sqrt{t}}{2}\frac{(-1)^k}{\xi_{k+1}}   L_k^{(2s)} (t/2),
 \end{array}
\right.     \label{g:L}
\end{equation}
with 
\begin{equation}
\xi_k=\xi_k(s)=\prod_{j=1}^k\left(s+\frac{j-1}{2}\right), \quad 
\xi_k^{\dag}=\xi_k(s^{\dag}), 
\end{equation}
where $L_k^{(r)}(x)$ are the Laguerre polynomials and $s^{\dag}=1/2-s$. 
Then, $\phi_{m,n}$ under the specialization (\ref{seed}) are given by 
\begin{equation}
\phi_{m,n}=N_{m,n}\det \left(g_{\lambda_j-j+i}^{(m+n+1-i)}\right)_{i,j=1}^{m+n},
\end{equation}
where the partition $\lambda$ and the normalization factor $N_{m,n}$ are given by 
\begin{equation}
\lambda=
\left\{
 \begin{array}{@{\,}ll}
  (3m-n-1,3m-n-4,\cdots,2n+5,2n+2,2n,2n,\cdots,4,4,2,2), \quad (m>n), \\
  (3n-m,3n-m-3,\cdots,2m+3,2m,2m,\cdots,4,4,2,2), \quad (m \le n),
 \end{array}
\right.
\end{equation}
and 
\begin{equation}
N_{m,n}=
\left\{
 \begin{array}{@{\,}ll}
 \displaystyle 
  (-1)^{n(n+1)/2} c_m d_n 
  \prod_{k=1}^n \hat{\zeta}_k \prod_{k=1}^m \zeta_k^{\dag} \prod_{k=1}^{m-n-1}\hat{\zeta}_k^{\dag}
   \quad (m>n), \\
 \displaystyle 
  (-1)^{n(n+1)/2} c_m d_n 
  \prod_{k=1}^n \hat{\zeta}_k \prod_{k=1}^m \zeta_k^{\dag} \prod_{k=1}^{n-m}  \zeta_k,
   \quad (m \le n), 
 \end{array}
\right.
\end{equation}
with 
\begin{equation}
\zeta_k=\prod_{j=1}^k(s+j-1), \quad \hat{\zeta}_k=\prod_{j=1}^k\left(s+\frac{2j-1}{2}\right),
\end{equation}
respectively. 
\end{prop}
This gives a different expression for the rational solutions
discussed in this paper. Studying the relationship between this formula
and our result might be an interesting problem.
\par\medskip

\noindent {\bf Acknowledgment}\quad The authors would like to thank
Prof. S. Okada for leading their attention to the universal characters.
They also thank Prof. M. Noumi, Prof. Y. Yamada and Prof. H. Watanabe
for useful suggestions and discussions.

\appendix
\section{Determinant Formula for the Umemura Polynomials \label{ume}}
In \cite{NY:P5}, Noumi and Yamada gave a determinant formula of
Jacobi-Trudi type for the Umemura polynomials in terms of 2-reduced Schur functions.
In this appendix, we give a brief review on this determinant
formula, and show that it is recovered as a special case of our formula.

We normalize the polynomials $U_{m,n}$ in Section 3 as
\begin{equation}
U_{m,n}=2^{m(m+1)/2+n(n+1)/2}T_{m,n}.
\end{equation}
Then we find that the functions $T_n=T_{0,n}(t,s)$ are monic polynomials 
generated by the Toda equation 
\begin{equation}
T_{n+1}T_{n-1}=
t\left[\left(\frac{d^2T_n}{dt^2}\right)T_n-\left(\frac{dT_n}{dt}\right)^2\right]
+\frac{dT_n}{dt}T_n
+\left(\frac{t}{8}-\frac{2s+1+3n}{4}\right)T_n^2,
\end{equation}
with initial conditions $T_{-1}=T_0=1$. The polynomials $T_n$ are called Umemura polynomials. 
It is easy to see that we have 
\begin{equation}
T_{-1,n}(t,s)=T_{0,n}(t,s+1/2). 
\end{equation}
Introducing $\hat{T}_n(t,s)$ by $\hat{T}_n(t,s)=T_{-n}(t,s)$, we have the following proposition. 
\begin{prop}
Let $\hat{T}_n=\hat{T}_n(t,s)$ be a sequence of polynomials in $t$ and $s$ defined through the Toda equation 
\begin{equation}
 \left.
 \begin{array}{@{\,}ll}
  \displaystyle 
   \hat{T}_{n+1}\hat{T}_{n-1}= 
   t\left[\left(\frac{d^2\hat{T}_n}{dt^2}\right)\hat{T}_n
          -\left(\frac{d\hat{T}_n}{dt}\right)^2\right]
   +\frac{d\hat{T}_n}{dt}\hat{T}_n
   +\left[\frac{t}{8}-\frac{1}{2}\left(s+\frac{1}{2}\right)+\frac{3}{4}n\right]\hat{T}_n^2,
 \end{array}
 \right.
\end{equation}
with initial conditions $\hat{T}_0=\hat{T}_1=1$. 
Then, the rational function
\begin{equation}
y=-\frac{\hat{T}_{n+1}(t,s)\hat{T}_n(t,s+1/2)}{\hat{T}_{n+1}(t,s+1)\hat{T}_n(t,s-1/2)},
\end{equation}
solves P$_{\rm V}$ with the parameters 
\begin{equation}
\kappa_{\infty}=s, \quad \kappa_0=s-n, \quad \theta=-n-1. 
\end{equation}
\end{prop}

The explicit formula for $\hat{T}_n$ was given by Noumi and Yamada,
which is expressed in terms of the 2-reduced Schur functions.
\begin{prop}\label{2-reduced}
Let $S_n=S_n(t_1,t_2,\cdots)$ for $n \ge 0$ be the Schur function
associated with a partition $\lambda=(n,n-1,\cdots,2,1)$.  Then, we have
\begin{equation}
\hat{T}_{n+1}(t,s)=N_n S_n 
\end{equation}
with 
\begin{equation}
 \left.
 \begin{array}{c}
  \displaystyle 
   N_n=2^{-n(n+1)}(2n-1)!!(2n-3)!!\cdots 3!!1!!, \\
  \displaystyle 
   t_j=\frac{t}{2}+\frac{-2s+n+1}{j}. 
 \end{array}
 \right.
\end{equation}
\end{prop}
It is easy to verify that Proposition \ref{2-reduced} is recovered by putting $m=0$ for the solutions (\ref{III}) with (\ref{para-III:1}) in Theorem \ref{main}.

\section{Proof of Lemmas \ref{dif:form:1} and \ref{dif:form:2} \label{PL}}
We first note that the following contiguity relations hold by definition,
\begin{equation}
p_k^{(r)}-p_{k-1}^{(r)}=p_k^{(r-1)}, \quad q_k^{(r)}-q_{k-1}^{(r)}=q_k^{(r-1)}, \label{rec:1}
\end{equation}
and 
\begin{equation}
(k+1)p_{k+1}^{(r)}=r p_k^{(r+1)}-x p_k^{(r+2)}, \quad 
(k+1)q_{k+1}^{(r)}=r q_k^{(r+1)}+x q_k^{(r+2)}. \label{rec:2}
\end{equation}

Let us prove Lemma \ref{dif:form:1}. 
Noticing that $p_1^{(r)}=1$ and $p_k^{(r)}=0$ for $k<0$, we see that
$R_{m,n}^{(r)}$ can be rewritten as 
\begin{equation}
R_{m,n}^{(r)}=
 \left|
  \begin{array}{cccccc}
   q_1^{(r)}        & q_0^{(r)}        & \cdots & q_{-m-n+3}^{(r)} & q_{-m-n+2}^{(r)} 
                    & q_{-m-n+1}^{(r)}  \\
   q_3^{(r)}        & q_2^{(r)}        & \cdots & q_{-m-n+5}^{(r)} & q_{-m-n+4}^{(r)} 
                    & q_{-m-n+3}^{(r)}  \\
   \vdots           & \vdots           & \ddots & \vdots           & \vdots 
                    & \vdots            \\
   q_{2m-1}^{(r)}   & q_{2m-2}^{(r)}   & \cdots & q_{m-n+1}^{(r)}  & q_{m-n}^{(r)}
                    & q_{m-n-1}^{(r)}   \\
   p_{n-m}^{(r)}    & p_{n-m+1}^{(r)}  & \cdots & p_{2n-2}^{(r)}   & p_{2n-1}^{(r)}
                    & p_{2n}^{(r)}      \\
   \vdots           & \vdots           & \ddots & \vdots           & \vdots
                    & \vdots            \\
   p_{-n-m+4}^{(r)} & p_{-n-m+5}^{(r)} & \cdots & p_2^{(r)}        & p_3^{(r)}
                    & p_4^{(r)}        \\
   p_{-n-m+2}^{(r)} & p_{-n-m+3}^{(r)} & \cdots & p_0^{(r)}        & p_1^{(r)}
                    & p_2^{(r)}        \\
   p_{-n-m}^{(r)}   & p_{-n-m+1}^{(r)} & \cdots & p_{-2}^{(r)}     & p_{-1}^{(r)}
                    & p_0^{(r)}        \\
  \end{array}
 \right|.    \label{edge:1}
\end{equation}
Subtracting the $(j-1)$-st column from the $j$-th column of
$R_{m,n}^{(r+1)}$ for $(j=m+n,m+n-1,\cdots,2)$ and using (\ref{rec:1}), we get
\begin{equation}
R_{m,n}^{(r+1)}=(-1)^m 
 \left|
  \begin{array}{cccccc}
   -q_1^{(r+1)}       & q_1^{(r)}        & \cdots & q_{-m-n+4}^{(r)} & q_{-m-n+3}^{(r)} \\
   -q_3^{(r+1)}       & q_3^{(r)}        & \cdots & q_{-m-n+6}^{(r)} & q_{-m-n+5}^{(r)} \\
   \vdots             & \vdots           & \ddots & \vdots           & \vdots           \\
   -q_{2m-1}^{(r+1)}  & q_{2m-1}^{(r)}   & \cdots & q_{m-n+2}^{(r)}  & q_{m-n+1}^{(r)}  \\
   p_{n-m}^{(r+1)}    & p_{n-m+1}^{(r)}  & \cdots & p_{2n-2}^{(r)}   & p_{2n-1}^{(r)}   \\
   \vdots             & \vdots           & \ddots & \vdots           & \vdots           \\
   p_{-n-m+4}^{(r+1)} & p_{-n-m+5}^{(r)} & \cdots & p_2^{(r)}        & p_3^{(r)}        \\
   p_{-n-m+2}^{(r+1)} & p_{-n-m+3}^{(r)} & \cdots & p_0^{(r)}        & p_1^{(r)}
  \end{array}
 \right|.    \label{shift:1}
\end{equation}
From (\ref{edge:1}) and (\ref{shift:1}), we obtain Lemma \ref{dif:form:1}. 

We next prove Lemma \ref{dif:form:2}.  Subtracting the $(i+1)$-st
column from the $i$-th column of $R_{m,n}^{(r)}$ for
$(i=1,2,\cdots,j,~j=m+n-1,m+n-2,\cdots,1)$ and using (\ref{rec:1}), we get
\begin{equation}
R_{m,n}^{(r)}=
 \left|
  \begin{array}{ccccc}
   q_1^{(r-m-n+1)}      & q_0^{(r-m-n+2)}      & \cdots  & q_{-m-n+3}^{(r-1)} & q_{-m-n+2}^{(r)} \\
   q_3^{(r-m-n+1)}      & q_2^{(r-m-n+2)}      & \cdots  & q_{-m-n+5}^{(r-1)} & q_{-m-n+4}^{(r)} \\
   \vdots               & \vdots               & \ddots  & \vdots             & \vdots           \\
   q_{2m-1}^{(r-m-n+1)} & q_{2m-2}^{(r-m-n+2)} & \cdots  & q_{m-n+1}^{(r-1)}  & q_{m-n}^{(r)}    \\
   (-1)^{m+n-1}p_{2n-1}^{(r-m-n+1)} & (-1)^{m+n-2}p_{2n-1}^{(r-m-n+2)}  & \cdots 
 & (-1)^1      p_{2n-1}^{(r-1)}     & (-1)^0      p_{2n-1}^{(r)}      \\
   \vdots           & \vdots           & \ddots           & \vdots           & \vdots \\
   (-1)^{m+n-1}p_3^{(r-m-n+1)}      & (-1)^{m+n-2}p_3^{(r-m-n+2)} & \cdots 
 & (-1)^1      p_3^{(r-1)}          & (-1)^0      p_3^{(r)}           \\
   (-1)^{m+n-1}p_1^{(r-m-n+1)}      & (-1)^{m+n-2}p_1^{(r-m-n+2)} & \cdots 
 & (-1)^1      p_1^{(r-1)}          & (-1)^0      p_1^{(r)}
  \end{array}
 \right|.    \label{shift:2}
\end{equation}
Noticing that $p_0^{(r)}=1$ and $p_k^{(r)}=0$ for $k<0$, we see that $R_{m,n}^{(r)}$ can be rewritten as 
\begin{equation}
R_{m,n}^{(r)}=
 \left|
  \begin{array}{ccccc}
   q_1^{(r-m-n)}      & q_0^{(r-m-n+1)}      & \cdots  & q_{-m-n+2}^{(r-1)} & q_{-m-n+1}^{(r)} \\
   q_3^{(r-m-n)}      & q_2^{(r-m-n+1)}      & \cdots  & q_{-m-n+4}^{(r-1)} & q_{-m-n+3}^{(r)} \\
   \vdots             & \vdots               & \ddots  & \vdots             & \vdots           \\
   q_{2m-1}^{(r-m-n)} & q_{2m-2}^{(r-m-n+1)} & \cdots  & q_{m-n}^{(r-1)}    & q_{m-n-1}^{(r)}  \\
   (-1)^{m+n}p_{2n}^{(r-m-n)} & (-1)^{m+n-1}p_{2n}^{(r-m-n+1)} & \cdots 
 & (-1)^1    p_{2n}^{(r-1)}   & (-1)^0      p_{2n}^{(r)}          \\
   \vdots             & \vdots               & \ddots  & \vdots             & \vdots \\
   (-1)^{m+n}p_2^{(r-m-n)}    & (-1)^{m+n-1}p_2^{(r-m-n+1)}    & \cdots 
 & (-1)^1    p_2^{(r-1)}      & (-1)^0      p_2^{(r)}             \\
   (-1)^{m+n}p_0^{(r-m-n)}    & (-1)^{m+n-1}p_0^{(r-m-n+1)}    & \cdots 
 & (-1)^1    p_0^{(r-1)}      & (-1)^0      p_0^{(r)}
  \end{array}
 \right|.    \label{edge:2}
\end{equation}
We add the $j$-th column multiplied by $(r-2+m+n-j)/x$ to the $(j+1)$-st
column of (\ref{edge:2}) for $(j=m+n,m+n-1,\cdots,1)$. Then using
(\ref{rec:1}) and (\ref{rec:2}), we obtain
\begin{eqnarray}
R_{m,n}^{(r)}&=&\prod_{j=0}^n(2j+1)\prod_{k=1}^m(r-m-n+2k-2)x^{-(m+n)} \nonumber \\
&\times&
 \left|
 \begin{array}{ccccc}
  \tilde{q}_1^{(r-m-n)}      & q_1^{(r-m-n-1)}      & \cdots
                             & q_{-m-n+3}^{(r-3)}   & q_{-m-n+2}^{(r-2)} \\
  \tilde{q}_3^{(r-m-n)}      & q_3^{(r-m-n-1)}      & \cdots
                             & q_{-m-n+5}^{(r-3)}   & q_{-m-n+4}^{(r-2)} \\
  \vdots                     & \vdots               & \ddots
                             & \vdots               & \vdots             \\
  \tilde{q}_{2m-1}^{(r-m-n)} & q_{2m-1}^{(r-m-n-1)} & \cdots
                             & q_{m-n+1}^{(r-3)}    & q_{m-n}^{(r-2)}    \\
  (-1)^{m+n}\hat{p}_{2n}^{(r-m-n)} & (-1)^{m+n}p_{2n+1}^{(r-m-n-1)} & \cdots
                                   & (-1)^2    p_{2n+1}^{(r-3)}     & (-1)^1 p_{2n+1}^{(r-2)} \\
  \vdots                           & \vdots                         & \ddots
                                   & \vdots                         & \vdots                  \\
  (-1)^{m+n}\hat{p}_2^{(r-m-n)}    & (-1)^{m+n}p_3^{(r-m-n-1)}      & \cdots
                                   & (-1)^2    p_3^{(r-3)}          & (-1)^1 p_3^{(r-2)}      \\
  (-1)^{m+n}\hat{p}_0^{(r-m-n)}    & (-1)^{m+n}p_1^{(r-m-n-1)}      & \cdots
                                   & (-1)^2    p_1^{(r-3)}          & (-1)^1 p_1^{(r-2)}
 \end{array}
 \right|.   \label{shift:3}
\end{eqnarray}
Lemma \ref{dif:form:2} follows from (\ref{shift:2}),(\ref{edge:2}) and (\ref{shift:3}).



\begin{thebibliography}{99}
\bibitem{JM}
Jimbo, M. and Miwa, T.:
Solitons and infinite dimensional Lie algebras. 
Publ. RIMS. Kyoto Univ. {\bf 19} 943-1001 (1983)
\bibitem{p2:rat}
Kajiwara, K. and Ohta, Y.:
Determinant structure of the rational solutions for the Painlev\'e II equation. 
J. Math. Phys. {\bf 37} 4693-4704 (1996)
\bibitem{p3:rat}
Kajiwara, K. and Masuda, T.:
On the Umemura polynomials for the Painlev\'e III equation. 
Phys. Lett. \textbf{A 260} 462-467 (1999)
\bibitem{p4:rat}
Kajiwara, K. and Ohta, Y.:
Determinant structure of the rational solutions for the Painlev\'e IV equation.
J. Phys. A: Math. Gen. {\bf 31} 2431-2446 (1998)
\bibitem{KLM}
Kitaev, A. V., Law, C. K. and McLeod, J. B.: 
Rational solutions of the fifth Painlev\'e equation. 
Differential and Integral Equations \textbf{7} 967-1000 (1994)
\bibitem{Koike}
Koike, K.:
On the decomposition of tensor products of the representations of the classical groups: by means of the universal characters. 
Adv. Math. {\bf 74} 57-86 (1989)
\bibitem{Um2}
Noumi, M., Okada, S., Okamoto, K. and Umemura, H.:
Special polynomials associated with the Painleve equations II.
In: Saito, M. H., Shimizu, Y., Ueno, K. (eds.) 
{\it Proceedings of the Taniguchi Symposium, 1997, Integrable Systems and Algebraic Geometry.}
Singapore: World Scientific, 1998, pp. 349-372
\bibitem{NY:P4}
Noumi, M and Yamada,Y.:
Symmetries in the fourth Painlev\'e equation and Okamoto polynomials. 
Nagoya Math. J. {\bf 153} 53-86 (1999)
\bibitem{NY:P5}
Noumi,M and Yamada,Y.:
Umemura polynomials for the Painlev\'e V equation. 
Phys. Lett. {\bf A247} 65-69 (1998)
\bibitem{NY1}
Noumi, M. and Yamada, Y.:
Higher order Painlev\'e equations of type $A_l^{(1)}$. 
Funkcial. Ekvac. {\bf 41} 483-503 (1998)
\bibitem{NY2}
Noumi, M and Yamada,Y.:
Affine Weyl groups, discrete dynamical systems and Painlev\'e equations. 
Commun. Math. Phys. {\bf 199} 281-295 (1998)
\bibitem{O1}
Okamoto, K.:
Studies on the Painlev\'e equations I, sixth Painlev\'e equation P$_{\rm VI}$. 
Annali di Matematica pura ed applicata {\bf CXLVI} 337-381 (1987)
\bibitem{O2}
Okamoto, K.:
Studies on the Painlev\'e equations II, fifth Painlev\'e equation P$_{\rm V}$. 
Japan J. Math. {\bf 13} 47-76 (1987)
\bibitem{O3}
Okamoto, K.:
Studies on the Painlev\'e equations III, second and fourth Painlev\'e equations, P$_{\rm II}$ and P$_{\rm IV}$. 
Math. Ann. {\bf 275} 222-254 (1986)
\bibitem{O4}
Okamoto, K.:
Studies on the Painlev\'e equations IV, third Painlev\'e equation P$_{\rm III}$.
Funkcial. Ekvac. {\bf 30} 305-332 (1987)
\bibitem{Tane}
Taneda, M.:
A proof of a conjecture associated with an algebraic solution of the sixth Painlev\'e equation. 
To appear in Nagoya Math. J.
\bibitem{Um1}
Umemura, H.:
Special polynomials associated with the Painlev\'e equations I. 
To appear in the Proceedings of the Workshop on ``Painlev\'e Transcendents", 
CRM, Montreal, Canada, 1996
\bibitem{Um:100}
Umemura, H.:
Irreducibility of the Painlev\'e equations--Evolution in the past 100 years. 
To appear in the Proceedings of the Workshop on ``Painlev\'e Transcendents", 
CRM, Montreal, Canada, 1996
\bibitem{YV}
Vorob'ev, A. P.: On rational solutions of the second Painlev\'e equation. 
Diff. Uravn. {\bf 1} 58-59 (1965)
\bibitem{Watanabe}
Watanabe, H.:
Solutions of the fifth Painlev\'e equation I. 
Hokkaido Math. J. {\bf 24} 231-267 (1995)
\bibitem{Ya}
Yamada, Y.:
Determinant formulas for the generalized Painlev\'e equations of type $A$.
Nagoya Math. J. {\bf 156} 123-134 (1999)
\end{thebibliography}
\end{document}